\documentclass[journal,onecolumn,12pt]{IEEEtran} 
\pdfoutput=1
\usepackage{amsmath}
\usepackage{amsthm}
\usepackage{amssymb}
\usepackage{bm}
\usepackage{xspace}
\usepackage{xcolor}
\usepackage{graphicx}
\usepackage{url}
\usepackage{framed}
\usepackage{float}
\usepackage{rotating}
\usepackage{verbatim}
\usepackage{listings}
\usepackage{lscape}

\usepackage{pgfplots}
\usepackage{pgf}
\usepackage{tikz}
\usetikzlibrary{arrows,shapes.misc,chains,scopes}
\usetikzlibrary{matrix}
\usetikzlibrary{calc}
\usetikzlibrary{fit}

\usepackage{pgfplotstable}
\usepackage{booktabs}
\usepackage{colortbl}

\newcommand{\executeiffilenewer}[3]{%
\ifnum\pdfstrcmp{\pdffilemoddate{#1}}%
{\pdffilemoddate{#2}}>0%
{\immediate\write18{#3}}\fi%
}

\graphicspath{{figures/}}

\theoremstyle{plain}
\newtheorem{lemma}{Lemma}

\newtheorem{theorem}{Theorem}

\newcounter{algocount}
\newcounter{examplecount}

\newcommand{\veca}{\boldsymbol{a}}
\newcommand{\vecb}{\boldsymbol{b}}
\newcommand{\vecB}{\boldsymbol{B}}

\newcommand{\vecu}{\boldsymbol{u}}
\newcommand{\vecU}{\boldsymbol{U}}

\newcommand{\setr}{\ensuremath{\mathbf{R}}\xspace}

\newcommand{\setb}{\ensuremath{\mathcal{B}}\xspace}
\newcommand{\setc}{\ensuremath{\mathcal{C}}\xspace}
\newcommand{\setd}{\ensuremath{\mathcal{D}}\xspace}

\newcommand{\sets}{\ensuremath{\mathcal{S}}\xspace}
\newcommand{\sett}{\ensuremath{\mathcal{T}}\xspace}

\newcommand{\setx}{\ensuremath{\mathcal{X}}\xspace}

\newcommand{\bmm}{\begin{matrix}}
\newcommand{\emm}{\end{matrix}}
\newcommand{\bpm}{\begin{pmatrix}}
\newcommand{\epm}{\end{pmatrix}}

\newcommand{\bsbm}{\left[\begin{smallmatrix}}
\newcommand{\esbm}{\end{smallmatrix}\right]}
\newcommand{\bspm}{\left(\begin{smallmatrix}}
\newcommand{\espm}{\end{smallmatrix}\right)}

\newcommand{\bbm}{\begin{bmatrix}}
\newcommand{\ebm}{\end{bmatrix}}

\DeclareMathOperator*{\argmax}{argmax}

\DeclareMathOperator{\expop}{\mathbb{E}}
\DeclareMathOperator{\entop}{\mathbb{H}}
\DeclareMathOperator{\miop}{\mathbb{I}}
\DeclareMathOperator{\kl}{\mathbb{D}}
\DeclareMathOperator{\rop}{\mathbb{R}}

\DeclareMathOperator*{\maximize}{maximize}
\DeclareMathOperator*{\st}{subject\;to}

\newcommand{\ogeq}[1]{\overset{\textnormal{(#1)}}{\geq}}

\usepackage{cite}
\usepackage{printlen}
\usepackage{siunitx}
\usepackage[normalem]{ulem}
\newcommand{\dentop}{\mathrm{h}}

\title{Achievable Rates for Shaped Bit-Metric Decoding}

\author{Georg B\"ocherer,~\IEEEmembership{Member,~IEEE}%
\thanks{A part of this work has been presented at ISIT 2014 in Honolulu \cite{bocherer2014probabilistic}.}%
\thanks{The author is with the Institute for Communications Engineering, Technical University of Munich, Munich D-80333, Germany (e-mail: georg.boecherer@tum.de).}%
}

\newcommand{\snr}{\ensuremath{\mathsf{SNR}}\xspace}
\newcommand{\power}{\mathsf{P}}

\newcommand{\tcr}[1]{\textcolor{black}{#1}}

\newcommand{\bits}{\boldsymbol{B}}
\newcommand{\cask}{\mathsf{C}}

\DeclareMathOperator{\supp}{supp}

\newcommand{\rate}{\mathsf{R}}
\newcommand{\gmi}{\rate_\textnormal{GMI}}
\newcommand{\rlm}{\rate_\textnormal{LM}}

\graphicspath{{figures/}}

\IEEEoverridecommandlockouts
\renewcommand{\rop}{\mathsf{R}}
\newcommand{\rbmd}{\rop_\textnormal{BMD}}

\pgfplotsset{
width=0.7\textwidth,
height=0.42\textheight
}
\newtheorem{remark}{Remark}

\begin{document}

\maketitle

\begin{abstract}
A new achievable rate for bit-metric decoding (BMD) is derived using random coding arguments. The rate expression can be evaluated for any input distribution, and in particular the bit-levels of binary input labels can be stochastically dependent. Probabilistic shaping with dependent bit-levels (shaped BMD), shaping of independent bit-levels (bit-shaped BMD) and uniformly distributed independent bit-levels (uniform BMD) are evaluated on the additive white Gaussian noise (AWGN) channel with Gray labeled bipolar amplitude shift keying (ASK). For 32-ASK at a rate of 3.8 bits/channel use, the gap to 32-ASK capacity is 0.008 dB for shaped BMD, 0.46 dB for bit-shaped BMD, and 1.42 dB for uniform BMD. These numerical results illustrate that dependence between the bit-levels is beneficial on the AWGN channel. The relation to \tcr{the LM rate and} the generalized mutual information (GMI) is discussed.
\end{abstract}


\section{Introduction}

Bit-interleaved coded modulation (BICM) combines higher order modulation with binary error correcting codes \cite{zehavi1992psk,caire1998bit}. This makes BICM attractive for practical application and BICM is widely used in standards, e.g., in DVB-T2/S2/C2. At a BICM receiver, \emph{bit-metric decoding} (BMD) is used \cite[Sec.~II]{martinez2009bit}.

For BMD, the channel input is labeled by bit strings of length $m$. The $m$ bit-levels are treated independently at the decoder. Let $\vecB=(B_1,B_2,\dotsc,B_m)$ denote a vector of $m$ binary random variables $B_i$, $i=1,2,\dotsc,m$, representing the bit-levels \tcr{with joint distribution $P_{\vecB}$ on $\{0,1\}^m$}. Consider the channel $p_{Y|\vecB}$ with output $Y$ \tcr{and input distribution $P_{\vecB}$. Define}
\begin{align}
\rbmd\tcr{(P_{\vecB})}:=\Bigl[\entop(\vecB)-\sum_{i=1}^m \entop(B_i|Y)\Bigr]^+\label{rate}
\end{align}
where $[\cdot]^+:=\max\{0,\cdot\}$, and where $\entop(\cdot)$ denotes entropy. \tcr{For independent bit-levels, we have $P_{\vecB}=\prod_{i=1}^m P_{B_i}$ and}
\begin{align}
\tcr{\textstyle\rbmd(\prod_{i=1}^m P_{B_i})}=\sum_{i=1}^m \miop(B_i;Y)\label{bicm}
\end{align}
\tcr{where $\miop(\cdot;\cdot)$ denotes mutual information}. Martinez \emph{et al.} showed in \cite{martinez2009bit} that \eqref{bicm} with independent and uniformly distributed bit-levels is achievable with BMD. We call this method \emph{uniform BMD}. \text{Guill\'en i F\`abregas} and Martinez \cite{ifabregas2010bit} generalized the result of \cite{martinez2009bit} to non-uniformly distributed independent bit-levels. We call this method \emph{bit-shaped BMD}. An important tool to assess the performance of decoding metrics is the \emph{generalized mutual information} (GMI) \cite[Sec.~2.4]{kaplan1993information}. An interpretation of uniform BMD and bit-shaped BMD as a GMI are given in \cite{martinez2009bit} and \cite{ifabregas2010bit}, respectively. In \cite[Sec.~4.2.4]{peng2012fundamentals}, the GMI is evaluated for a bit-metric. It is observed that the GMI increases when the bits are dependent. We call this approach \emph{shaped GMI}. \tcr{Another method to evaluate decoding metrics is the LM rate, see \cite{ganti2000mismatched} and references therein}. The LM rate is applied to BICM in \cite{peng2013improved}.

\begin{figure*}
%\footnotesize
\centering
\includegraphics[width=0.7\textwidth]{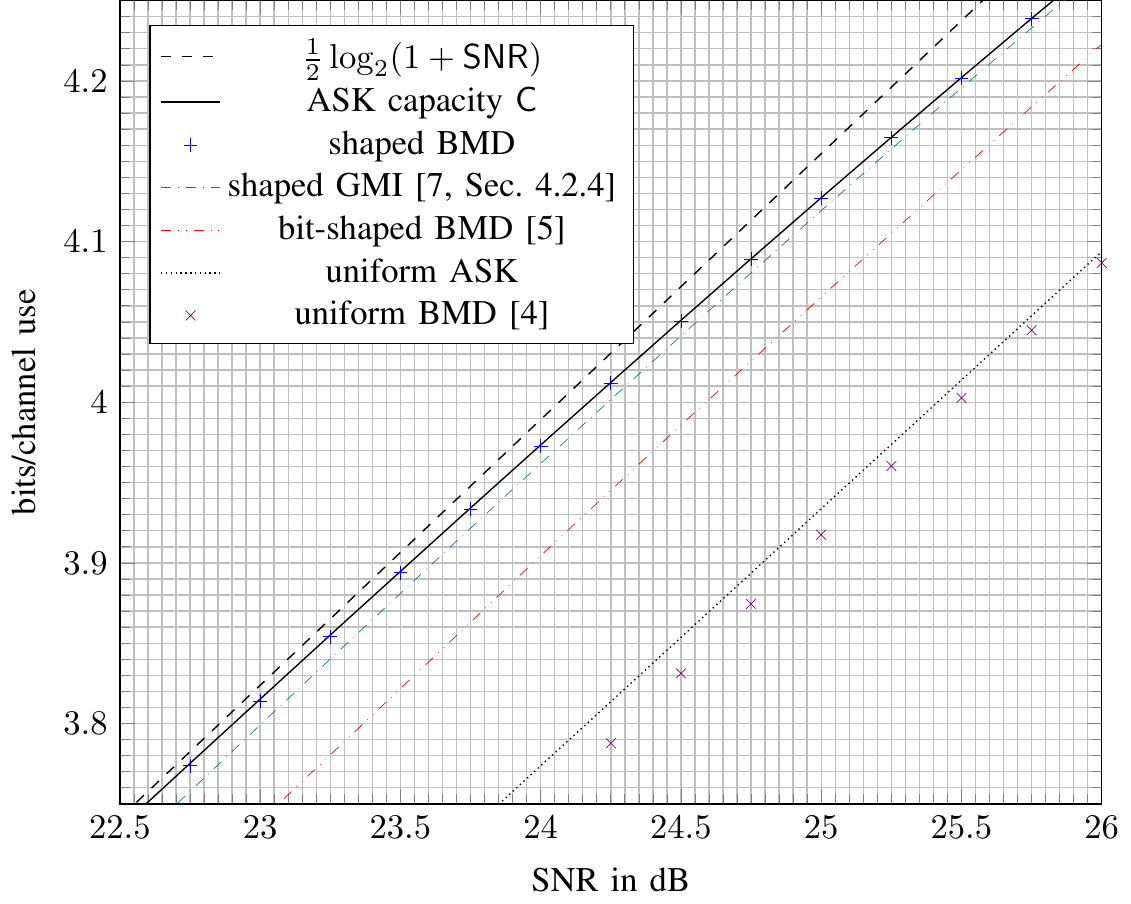}
\caption{\normalsize Achievable rates for bipolar ASK with 32 equidistant signal points, see Sec.~\ref{sec:awgn}. At 3.8 bits/channel use, bit-shaped BMD is \SI{0.46}{\decibel} less energy efficient than shaped BMD.}
\label{fig:gmi}
\end{figure*}

Our main contribution is to show that $\rbmd$ in \eqref{rate} with arbitrarily distributed bit-levels is achievable with BMD. In particular, the bit-levels can be dependent, in which case $\rbmd$ is not equal to \eqref{bicm}. We call our method \emph{shaped BMD}. For example, consider the additive white Gaussian noise (AWGN) channel with bipolar amplitude shift keying (ASK), see Sec.~\ref{sec:awgn} for details. We display information rate results for 32-ASK in Fig.~\ref{fig:gmi}. At a rate of 3.8 bits/channel use, the gap to ASK capacity of shaped BMD is \SI{0.008}{\decibel}, the gap for shaped GMI is \SI{0.1}{\decibel}, the gap for bit-shaped BMD is \SI{0.46}{\decibel}, and the gap is \SI{1.42}{\decibel} for uniform BMD. Dependence between the bit-levels is thus beneficial on the AWGN channel. The rate expression \eqref{rate} is used in \cite{steiner2016protograph} to construct surrogate channels, which are used to design low-density parity-check codes for shaped BMD. In \cite{buchali2016rate}, $\rbmd$ is used to estimate achievable rates in fiber-optic transmission experiments.

This paper is organized as follows. We state our main result in Sec.~\ref{sec:rate}. \tcr{We relate $\rbmd$ to the LM rate and the GMI in Sec.~\ref{sec:gmi}} and we discuss its application to the AWGN channel in Sec.~\ref{sec:awgn}. Sec.~\ref{sec:conclusions} concludes the paper and the appendix provides technical results.

\section{\tcr{Main Result}}
\label{sec:rate}

Let \tcr{$p_{Y|\vecB}$ be a memoryless channel with input $\vecB=(B_1,B_2,\dotsc,B_m)$.} Let $\mathcal{C}$ be a codebook with code words $\vecb^n=(\vecb_1,\vecb_2,\dotsc,\vecb_n)$ with $\vecb_i\in\{0,1\}^m$. We denote the $i$th bit-level of a code word by $b_i^n=(b_{i1},b_{i2},\dotsc,b_{in})$. A \emph{bit-metric decoder} uses a decision rule of the form
\begin{align}
\argmax_{\vecb^n\in\setc} \prod_{i=1}^m q_i(y^n,b_i^n)\label{eq:qmetric}
\end{align}
\tcr{where $y^n$ is the decoder's observation of the channel output and }where for each bit-level $i$, the value of the bit-metric $q_i(y^n,b_i^n)$ depends on $P_{\vecB}p_{Y|\vecB}$ only via the marginal
\begin{align}
P_{B_i}(b)p_{Y|B_i}(y|b)=\sum_{\veca\in\{0,1\}^m\colon a_i=b}p_{Y|\vecB}(y|\veca)P_{\vecB}(\veca).\label{eq:pby}
\end{align}
\begin{theorem}\label{theo:ssbcm}
Let $P_{\vecB}$ be a distribution on $\{0,1\}^m$ and let \tcr{$p_{Y|\vecB}$ be a memoryless channel}. The rate $\rbmd\tcr{(P_{\vecB})}$ can be achieved by a bit-metric decoder.
\end{theorem}
\begin{IEEEproof}
\tcr{The theorem follows by Lemma~\ref{lem:lm}, see Sec.~\ref{sec:gmi}. For channels with finite output alphabets, we give a proof by typicality in Appendix~\ref{app:bmdrate}}.
\end{IEEEproof}
\subsection{Dependent Bit-Levels Can Be Better}
We develop a simple contrived example to show that dependent bit-levels can be better than independent bit-levels. Consider the identity channel with input label $B_1B_2$ and transition probabilities
\begin{align*}
P_{Y|B_1B_2}(ab|ab)=1,\quad\forall ab\in\{00,01,10,11\}.
\end{align*}
Consider the input cost function $f$ satisfying
\begin{align*}
f(00)=f(11)=\infty,\;f(01)=f(10)=0
\end{align*}
and suppose we impose the average cost constraint $\expop[f(B_1B_2)]<\infty$, where $\expop[\cdot]$ denotes expectation. For independent bit-levels $B_1$ and $B_2$, this constraint can be achieved only by $P_{B_1}(0)=P_{B_2}(1)=1$ or  $P_{B_1}(1)=P_{B_2}(0)=1$. In both cases, we have
\begin{align}
\entop(B_1B_2)-\sum_{i=1}^2 \entop(B_i|Y)=0=\rbmd\tcr{(P_{B_1}P_{B_2})}.
\end{align}
We next choose $P_{B_1B_2}(01)=P_{B_1B_2}(10)=1/2$, which makes the bit-levels dependent. The average input cost is zero and we have
\begin{align}
\entop(B_1B_2)-\sum_{i=1}^2 \entop(B_i|Y)=1=\rbmd\tcr{(P_{B_1B_2})}.
\end{align}
We conclude that for the considered input-constraint channel, no positive rate is achievable with independent bit-levels and any rate below one is achievable with dependent bit-levels.

\subsection{$\entop(\vecB)-\sum_{i=1}^m\entop(B_i|Y)$ Can Be Negative}
\label{subsec:negative}
Consider the erase-all channel with output alphabet $\{e\}$ and transition probabilities 
\begin{align*}
P_{Y|B_1B_2}(e|ab)=1,\quad\forall ab\in\{00,01,10,11\}.
\end{align*}
For the input distribution $P_{B_1B_2}(01)=P_{B_1B_2}(10)=1/2$, we compute
\begin{align}
\entop(B_1B_2)-\sum_{i=1}^2 \entop(B_i|Y)=1-2=-1.
\end{align}
Thus, $\rbmd\tcr{(P_{B_1B_2})}=[-1]^+=0$.

%\color{blue}
  \section{$\rbmd$, LM Rate, and GMI}
\label{sec:gmi}

\subsection{Random Coding}
\label{subsec:rand}
Consider a memoryless channel $p_{Y|\vecB}$ with input alphabet $\{0,1\}^m$ and real-valued output $Y$. Let $\setc\subseteq\{0,1\}^{mn}$ be a codebook with block length $n$ over the alphabet $\{0,1\}^m$ of size $|\setc|=2^{nR}$. The decision rule of a \emph{maximum likelihood} (ML) decoder is
\begin{align}
\hat{\vecb}^n=\argmax_{\vecb^n\in\setc}\prod_{i=1}^n p_{Y|\vecB}(y_i|\vecb_i)
\end{align}
where $y^n$ is the decoder's observation of the channel output. If more than one code word maximizes the likelihood function, the $\argmax$ operator selects one maximizing code word randomly. For the ensemble of codes whose code words are drawn iid according to $P_{\vecB}^n$, the following holds (see, e.g., \cite[Chap.~7]{gallager1968information}): the average probability of erroneous ML-decoding (the average is both over the codes in the ensemble and the code words) approaches zero for $n$ approaching infinity if
\begin{align}
R<\miop(\vecB;Y)
\end{align}
which shows that $\miop(\vecB;Y)$ is an achievable rate for ML-decoding. A \emph{mismatched decoder} \cite[Sec.~II]{ganti2000mismatched} uses a metric $q\colon \{0,1\}^m\times\setr\to\setr$ instead of $p_{Y|\vecB}$ and the mismatched decoding rule is
\begin{align}
\hat{\vecb}^n=\argmax_{\vecb^n\in\setc}\prod_{i=1}^n q(y_i,\vecb_i).\label{eq:mmd}
\end{align}

\subsection{LM Rate and $\rbmd$}
Let $s\geq 0$ be a non-negative scalar and let $r\colon\{0,1\}^m\to\setr$ be a real-valued function defined on the input alphabet $\{0,1\}^m$. Define
\begin{align}
\rate(P_{\vecB},q,s,r)=\expop\left[\log_2\frac{q(Y,\vecB)^s r(\vecB)}{\sum_{\vecb\in\supp P_{\vecB}}P_{\vecB}(\vecb)q(Y,\vecb)^s r(\vecb)}\right]
\end{align}
where $\supp P_{\vecB}:=\{\vecb\in\{0,1\}^m\colon P_{\vecB}(\vecb)>0\}$ is the support of $P_{\vecB}$. By \cite[Theorem~2]{ganti2000mismatched}, the LM rate
\begin{align}
\rlm(P_X,q)=\max_{s\geq 0,r}\rate(P_{\vecB},q,s,r)\label{eq:lmrate}
\end{align}
is achievable by the mismatched decoder \eqref{eq:mmd}. This implies that for each $s\geq 0$ and function $r$, the rate $[\rate(P_{\vecB},q,s,r)]^+$ is also an achievable rate for the mismatched decoder. The next lemma states that $\rbmd$ is an instance of $[\rate(P_{\vecB},q,s,r)]^+$ for a particular choice of $q,s,r$.
\begin{lemma}\label{lem:lm} Consider the memoryless channel $p_{Y|\vecB}$ with input distribution $P_{\vecB}$ and define
\begin{align}
q_\textnormal{BMD}(y,\vecb)&=\prod_{i=1}^m p_{Y|B_i}(y|b_i)\\
s_\textnormal{BMD}&=1\\
r_\textnormal{BMD}(\vecb)&=\frac{\prod_{i=1}^m P_{B_i}(b_i)}{P_{\vecB}(\vecb)}.
\end{align}
\begin{enumerate}
\item We have
\begin{align}
\rbmd(P_{\vecB})\leq[\rate(P_{\vecB},q_\textnormal{BMD},s_\textnormal{BMD},r_\textnormal{BMD})]^+
\end{align}
with equality if and only if $P_{\vecB}$ is strictly positive, i.e., if $P_{\vecB}(\vecb)>0$ for all $\vecb\in\{0,1\}^m$. 
\item  $\rbmd$ is an achievable rate for the mismatched decoder \eqref{eq:mmd} for $q=q_\textnormal{BMD}$.
\item $\rbmd$ is less or equal to the LM rate, i.e.,
\begin{align}
\rbmd(P_{\vecB})\leq \rlm(P_{\vecB},q_\textnormal{BMD}).
\end{align}
\end{enumerate}
\end{lemma}
\begin{IEEEproof}
We prove statement 1) in Appendix~\ref{app:gmi}. Statements 2) and 3) now follow by 1) and \eqref{eq:lmrate}. 
\end{IEEEproof}
\begin{remark}\label{remark:code mismatch}
The achievability of $\rbmd$ for channels with finite output alphabets is show in Appendix~\ref{app:bmdrate} by the following code construction, which is different from the construction in Sec.~\ref{subsec:rand}:
\begin{itemize}
\item The codebook $\setc$ is generated iid according to a uniform distribution on the alphabet $\{0,1\}^m$.
\item The transmitter only transmits code words that are $P_{\vecB}$ typical (for the formal definition of typicality, see Appendix~\ref{app:typical}), which results in the signaling set 
\begin{align}
\sets=\{\vecb^n\in\setc\colon\vecb^n \text{ is $P_{\vecB}$ typical}\}.
\end{align}
\item The receiver uses a bit-wise typicality decoder, i.e., it outputs $\vecb^n$, if it is the only code word whose bit levels $b_i^n$ are jointly $P_{B_iY}$ typical with the observed channel output $y^n$, for $i=1,2,\dotsc,m$. Thus, the decoder evaluates its decoding metric for the decoding set
\begin{align}
\setd=\{\vecb^n\in\setc\colon b_i^n \text{ is $P_{B_i}$ typical, for }i = 1,2,\dotsc,m\}.
\end{align}
\end{itemize}
In the case of dependent bit-levels, the signaling set $\sets$ and the decoding set $\setd$ are not equal. This \emph{codebook mismatch} is also present in the practical implementation of shaped BMD \cite{bocherer2015bandwidth}. A related work on codebook mismatch is \cite{achtenberg2013theoretic}.  
\end{remark}

\subsection{Generalized Mutual Information}

By setting $r(\cdot)=1$ and maximizing over $s$, we obtain the GMI
\begin{align}
\gmi(P_{\vecB},q)=\max_{s\geq 0}\rate(P_{\vecB},q,s,1).
\end{align}
The achievability of $\gmi$ by the mismatched decoder \eqref{eq:mmd} now follows by observing that $\gmi(P_{\vecB},q)\leq\rlm(P_{\vecB},q)$ or by invoking \cite[Sec.~2.4]{kaplan1993information}. In \cite[Sec.~4.2.4]{peng2012fundamentals}, dependent bit-levels are used together with the metric
$q_\textnormal{BMD}$, which yields the shaped GMI rate
\begin{align}
\mathsf{R}_\textnormal{sGMI}(P_{\vecB}):=\gmi(P_{\vecB},q_\textnormal{BMD}).
\end{align}
In the next section, we will see that for bipolar ASK on the AWGN channel, $\rbmd$ is larger than $\mathsf{R}_\textnormal{sGMI}$ .

\color{black}

\section{$2^m$-ASK Modulation for the AWGN Channel}
\label{sec:awgn}
The signal constellation of bipolar ASK is given by
\begin{align}
\setx_\text{ASK}=\{\pm 1,\pm 3,\dotsc,\pm(2^m-1)\}.
\end{align}
The points $x\in\setx_\text{ASK}$ are labeled by a binary vector $\vecB=(B_1,\dotsc,B_m)$. We use the \emph{Binary Reflected Gray Code} (BRGC) \cite{gray1953pulse}. The labeling influences the rate that is achievable by BMD, see, e.g., \cite[Sec.~VI.C]{bocherer2015bandwidth}. To control the transmit power, the channel input $x_{\vecB}$ is scaled by a positive real number $\Delta$. The input-output relation of the AWGN channel is
\begin{align}
Y=\Delta\cdot x_{\vecB}+Z
\end{align}
where $Z$ is zero mean Gaussian noise with variance one. The input is subject to an average power constraint $\power$, i.e., $\Delta$ and $P_{\vecB}$ must satisfy $\expop[(\Delta x_{\vecB})^2]\leq\power$. The ASK capacity is 
\begin{align}
\cask=\max_{\Delta,P_{\vecB}\colon\expop[(\Delta x_{\vecB})^2]\leq\power}\miop(\vecB;Y).\label{cm} 
\end{align}
The optimal parameters $\Delta^*,P_{\vecB}^*$ can be calculated using the Blahut-Arimoto algorithm \cite{blahut1972computation,arimoto1972algorithm} and they can be approximated closely by maximizing over the family of Maxwell-Boltzmann distributions \cite{kschischang1993optimal}, see also \text{\cite[Sec.~III]{bocherer2015bandwidth}}. We evaluate $\rbmd$ (shaped BMD) and $\mathsf{R}_\text{sGMI}$ (shaped GMI) in $\Delta^*,P_{\vecB}^*$. In Fig.~\ref{fig:gmi}, we plot for 32 signal points ($m=5$) the ASK capacity $\cask$ and the information rate curves of shaped BMD and shaped GMI together with the corresponding rate curves that result from uniform inputs. Since we normalized the noise power to one, the signal-to-noise ratio (SNR) in dB is given by
\begin{align}
\snr=10\log_{10}\frac{\expop[(\Delta x_{\vecB})^2]}{1}.
\end{align}
The gap between the 32-ASK capacity $\cask$ and the shaped BMD rate $\rbmd$ is negligibly small over the considered SNR range. At 3.8 bits/channel use, the gap between $\cask$ and $\rbmd$ is \SI{0.008}{\decibel} and the gap of sGMI is $\SI{0.1}{\decibel}$. For comparison, we calculate the bit-shaped BMD rate. The optimization problem is
\begin{align}
\begin{split}
\maximize_{P_{\bits},\Delta}\quad&\sum_{i=1}^m\miop(B_i;Y)\\
\st\quad &P_{\bits} = \prod_{i=1}^{m}P_{B_i},\quad
\expop[(\Delta x_{\bits})^2]\leq\power.
\end{split}
\end{align}
This is a non-convex optimization problem \cite{alvarado2011high,bocherer2012efficient} so we calculate a solution by exhaustive search over the bit distributions with a precision of $\pm 0.005$. The resulting rate curve is displayed in Fig.~\ref{fig:gmi}. We observe that bit-shaped BMD (independent bit-levels) is \SI{0.46}{\decibel} less energy efficient than shaped BMD (dependent bit-levels) at 3.8 bits/channel use.

\section{Conclusions}
\label{sec:conclusions}
The achievable rate in \eqref{rate} allows dependence between the bit-levels while the achievable rate in \eqref{bicm} (see \cite{martinez2009bit},\cite{ifabregas2010bit}) requires independent bit-levels. We have shown that on the AWGN channel under bit-metric decoding, dependent bit-levels can achieve higher rates than independent bit-levels.

Interesting research directions are to study codebook mismatch (see Remark~\ref{remark:code mismatch}) and to study error exponents for shaped BMD.

\section*{Acknowledgment}

This work was supported by the German Federal Ministry of Education and Research in the framework of an Alexander von Humboldt Professorship.

\appendices
\section{Typical Sequences}
\label{app:typical}
We use letter-typical sequences as defined in \cite[Sec.~1.3]{kramer2007topics}. Consider a \emph{discrete memoryless source} (DMS) $P_X$ with a finite alphabet $\setx$. For $x^n\in\setx^n$, let $N(a|x^n)$ be the number of times that letter $a\in\setx$ occurs in $x^n$, i.e.,
\begin{align}
N(a|x^n)=|\{i\colon x_i=a\}|.
\end{align}
We say $x^n$ is $\epsilon$-letter-typical with respect to $P_X$ if for each letter $a\in\setx$,
\begin{align}
(1-\epsilon)P_X(a)\leq\frac{N(a|x^n)}{n}\leq (1+\epsilon)P_X(a),\quad\forall a\in\setx.\label{eq:def:typ}
\end{align}
Let $\sett_\epsilon^n(P_X)$ be the set of all sequences $x^n$ that fulfill \eqref{eq:def:typ}. The sequences \eqref{eq:def:typ} are called typical in \cite[Sec.~3.3]{masseyapplied1},\cite[Sec.~2.4]{elgamal2011network} and robust typical in \cite[Appendix]{orlitsky2001coding}. We next list the properties of typical sequences that we need in this work and whenever possible, we refer for the proofs to the literature. Define
\begin{align}
\mu_X:=\min_{a\in\supp P_X}P_X(a).\label{eq:def:mu}
\end{align}
\begin{lemma}[{Typicality, \cite[Theorem~1.1]{kramer2007topics},\cite[Lemma~19]{orlitsky2001coding}}]
Suppose $0<\epsilon<\mu_X$. We have
\begin{align}
(1-\delta_\epsilon(n,P_X))2^{n(1-\epsilon)\entop(X)}\leq |\sett_\epsilon^n(P_X)|\label{eq:typ 2}
\end{align}
where $\delta_\epsilon(P_X,n)$ is such that $\delta_\epsilon(P_X,n)\overset{n\to\infty}{\to}$ exponentially fast in $n$.
\end{lemma}
The definitions and properties of typicality apply in particular when the random variable $X$ stands for a tuple of random variables, e.g., $X=(Y,Z)$. If $x^n=(y^n,z^n)$ is typical, then $y^n$ and $z^n$ are called jointly typical. 
\begin{lemma}[{Marginal Typicality, \cite[Lemma~21]{orlitsky2001coding},\cite[Sec.~1.5]{kramer2007topics}}]\label{lem:marginal}
Joint typicality implies marginal typicality, i.e., $\sett_\epsilon^n(P_{YZ})\subseteq\sett_\epsilon^n(P_Y)\times\sett_\epsilon^n(P_Z)$.
\end{lemma}

\begin{lemma}[Mismatched Typicality]
Suppose $\epsilon>0$, $X^n$ is emitted by the DMS $P_X$ and $\supp P_{\tilde{X}}\subseteq \supp P_X$. We have
\begin{align}
(1-\delta_\epsilon(P_{\tilde{X}},n))2^{-n[\kl(P_{\tilde{X}}\Vert P_X)-\epsilon\log_2(\mu_{\tilde{X}}\mu_X)]}\leq \Pr[X^n\in\sett_\epsilon^n(P_{\tilde{X}})].\label{eq:mis}
\end{align}
\end{lemma}
\begin{IEEEproof}
For $x^n\in\sett_\epsilon^n(P_{\tilde{X}})$, we have
\begin{align}
P_X^n(x^n)&=\prod_{a\in\supp P_{\tilde{X}}}P_X(a)^{N(a|x^n)}\nonumber\\
&\geq \prod_{a\in\supp P_{\tilde{X}}}P_X(a)^{n(1+\epsilon)P_{\tilde{X}}(a)}\nonumber\\
&= 2^{\sum_{a\in\supp P_{\tilde{X}}} n(1+\epsilon)P_{\tilde{X}}(a)\log_2 P_X(a)}.\label{eq:prebound mis}
\end{align}
Now, we have
\begin{align}
\Pr[X^n\in \sett_\epsilon^n(P_{\tilde{X}})]&=\sum_{x^n\in\sett_\epsilon^n(P_{\tilde{X}})}P_X^n(x^n)\\
&\ogeq{\ref{eq:prebound mis}}\sum_{x^n\in\sett_\epsilon^n(P_{\tilde{X}})}2^{\sum_{a\in\supp P_{\tilde{X}}} n(1+\epsilon)P_{\tilde{X}}(a)\log_2 P_X(a)}\\
&\ogeq{\ref{eq:typ 2}}(1-\delta_\epsilon(n,P_{\tilde{X}}))2^{n(1-\epsilon)\entop(\tilde{X})}2^{\sum_{a\in\supp P_{\tilde{X}}} n(1+\epsilon)P_{\tilde{X}}(a)\log_2 P_X(a)}\\
&=(1-\delta_\epsilon(n,P_{\tilde{X}}))2^{-n[\kl(P_{\tilde{X}}\Vert P_X)-\epsilon\entop(\tilde{X})+\epsilon\sum_{a\in\supp P_{\tilde{X}}} P_{\tilde{X}}(a)\log_2 P_X(a)]}\\
&\geq (1-\delta_\epsilon(n,P_{\tilde{X}}))2^{-n[\kl(P_{\tilde{X}}\Vert P_X)+\epsilon\log_2(\mu_{\tilde{X}}\mu_X)]}.
\end{align}
\end{IEEEproof}
Define now the set
\begin{align}
\sett_\epsilon^n(P_{XY}|x^n):=\Bigl\{y^n\colon (x^n,y^n)\in\sett_\epsilon^n(P_{XY})\Bigr\}.
\end{align}
Note that $\sett_\epsilon^n(P_{XY}|x^n)=\emptyset$ if $x^n\notin\sett_\epsilon^n(P_X)$.
\begin{lemma}[{Conditional Typicality, \cite[Theorem~1.2]{kramer2007topics},\cite[Lemma~22, 24]{orlitsky2001coding}}]
Suppose $0< \epsilon_1<\epsilon_2<\mu_{XY}$. For any $x^n\in\setx^n$, we have
\begin{align}
|\sett_{\epsilon_2}^n(P_{XY}|x^n)|\leq 2^{n(1+\epsilon_2)\entop(Y|X)}.\label{eq:cond 2}
\end{align}
Suppose $x^n\in\sett_{\epsilon_1}^n(P_X)$ and that $(X^n,Y^n)$ is emitted by the DMS $P_{XY}$. We have
\begin{align}
1-\delta_{\epsilon_1,\epsilon_2}(P_{XY},n)\leq \Pr[Y^n\in\sett_{\epsilon_2}^n(P_{XY}|x^n)|X^n=x^n]\label{eq:cond 3}
\end{align}
where $\delta_{\epsilon_1,\epsilon_2}(P_{XY},n)$ approaches zero exponentially fast in $n$.
\end{lemma}

\section{Proof of Theorem \ref{theo:ssbcm}}
\label{app:bmdrate}

We prove Theorem~\ref{theo:ssbcm} by random coding arguments. In the following, let $0<\epsilon_1<\epsilon_2<\mu_{\vecB Y}$, \tcr{where by \eqref{eq:def:mu}, $\displaystyle\mu_{\vecB Y}=\min_{\vecb,y\colon P_{\vecB Y}(\vecb y)>0}P_{\vecB Y}(\vecb y)$.}

\emph{Code Construction:}
%%%
Choose $2^{n(R+\tilde{R})}$ code words $\vecU^n(w,v)$, $w=1,2,\dotsc,2^{nR}$, $v=1,2,\dotsc,2^{n\tilde{R}}$ of length $n$ by choosing the $n\cdot 2^{n(R+\tilde{R})}$ symbols independent and uniformly distributed according to the uniform distribution $P_{\vecU}$ on $\{0,1\}^m$. Let $\setc$ be the resulting codebook. 
%%%

\emph{Encoding:} Given message $w\in\{1,2,3,\dotsc,2^{nR}\}$, try to find a $v$ such that $\vecU^n(w,v)\in\sett^n_{\epsilon_1}(P_{\vecB})$. If there is such a $v$, transmit $\vecU^n(w,v)$. If there is no such $v$, declare an error.

\emph{Decoding:} We define the bit-metric
\begin{align}
q_i(y^n,b_i^n)=\begin{cases}
1,&(b_i^n,y^n)\in\sett^n_{\epsilon_2}(P_{B_iY})\\
0,&\text{otherwise}.
\end{cases}
\end{align}
The corresponding decoding metric is
\begin{align}
q(y^n,\vecb^n)=\prod_{i=1}^m q_i(y^n,b_i^n).
\end{align}
Define the set
$\hat{\setb}(y^n):=\{\vecb^n\in\setc\colon q(y^n,\vecb^n)=1\}$. The decoder output is
\begin{align}
\begin{cases}
\vecb^n,&\text{if }\hat{\setb}(y^n)=\{\vecb^n\}\\
\text{error},&\text{otherwise}.
\end{cases}
\end{align}

\emph{Analysis:} Suppose message $w$ should be transmitted. The first error event is
\begin{align}
&\mathcal{E}_1:=\bigl\{\nexists v\colon\vecU^n(w,v)\in\sett_{\epsilon_1}^n(P_{\vecB})\bigl\}.
\end{align}
Suppose now $\mathcal{E}_1$ did not occur and for some $v$, $\vecU^n(w,v)=\vecb^n$ with $\vecb^n\in\sett_{\epsilon_1}^n(P_{\vecB})$. The vector $\vecb^n$ is transmitted. The second error event can occur at the decoder, and it is given by
\begin{align}
&\mathcal{E}_2:=\bigl\{\vecb^n\notin\hat{\setb}(Y^n)|\vecU^n(w,v)=\vecb^n\bigl\}.
\end{align}
Suppose next that the second error event did not occur. This implies in particular that $Y^n\in\sett_{\epsilon_2}^n(P_Y)$. Suppose that $Y^n=y^n$ for some $y^n\in\sett_{\epsilon_2}^n(P_Y)$. The third error event is now
\begin{align}
&\mathcal{E}_3:=\bigl\{\exists\tilde{w},\tilde{v}\colon \tilde{w}\neq w \text{ and }\vecU^n(\tilde{w},\tilde{v})\in\hat{\setb}(y^n)|Y^n=y^n\bigr\}.
\end{align}

\emph{First error event:}
By \eqref{eq:mis}, we have
\begin{align}
\Pr[\vecU^n\in\mathcal{T}_{\epsilon_1}^n(P_{\vecB})]\geq[1-\delta_{\epsilon_1}(P_{\vecB},n)]2^{-n[D(P_{\vecB}\Vert P_{\vecU})-\epsilon_1\log_2(\mu_{\vecB}\mu_{\vecU})]}
\end{align}
\tcr{where by  \eqref{eq:def:mu}, $\displaystyle\mu_{\vecB}=\min_{\vecb\colon P_{\vecB}(\vecb)>0}P_{\vecB}(\vecb)$ and $\displaystyle\mu_{\vecU}=\min_{\vecu\colon P_{\vecU}(\vecu)>0}P_{\vecU}(\vecu)$. Note that since $P_{\vecU}$ is uniform on $\{0,1\}^m$, we have $\mu_{\vecU}=2^{-m}$}. For large enough $n$, we have $\delta_{\epsilon_1}(P_{\vecB},n)\leq 1/2$. The probability to generate $2^{n\tilde{R}}$ sequences $\vecU^n(w,v)$, $v=1,2,\dotsc,2^{n\tilde{R}}$, that are \emph{not} in $\mathcal{T}_{\epsilon_1}^n(P_{\vecB})$ is thus bounded from above by
\begin{align}
\Bigl(1-\frac{1}{2}2^{-n[D(P_{\vecB}\Vert P_{\vecU})-\epsilon_1\log_2(\mu_{\vecB}\mu_{\vecU})]}\Bigr)^{2^{n\tilde{R}}}\leq\exp\left[-\frac{1}{2}2^{-n[D(P_{\vecB}\Vert P_{\vecU})-\epsilon_1\log_2(\mu_{\vecB}\mu_{\vecU})]}2^{n\tilde{R}}\right]\label{eq:rate:ebound1}
\end{align} 
where \tcr{inequality in \eqref{eq:rate:ebound1}} follows by $(1-r)^s\leq\exp(-rs)$. This probability tends to zero if
\begin{align}
\tilde{R}>D(P_{\vecB}\Vert P_{\vecU})+\epsilon_1\log_2\frac{1}{\mu_{\vecB}\mu_{\vecU}}.\label{eq:rate:cbmm}
\end{align}
We conclude that if
\begin{align}
\tilde{R}>D(P_{\vecB}\Vert P_{\vecU})\label{eq:rate}
\end{align}
then for small enough positive $\epsilon_1$ and large enough $n$, we have $\vecU^n(w,v)\in\sett_{\epsilon_1}^n(P_{\vecB})$ for some $v\in\{1,2,\dotsc,2^{n\tilde{R}}\}$ with high probability.

\emph{Second error event:} By \eqref{eq:cond 3}, the probability 
\begin{align*}
\Pr[(\vecb^n,Y^n)\in\sett_{\epsilon_2}^n(P_{\vecB Y})|\vecU^n(w,v)=\vecb^n]=\Pr[Y^n\in\sett_{\epsilon_2}^n(P_{\vecB Y}|\vecb^n)|\vecU^n(w,v)=\vecb^n]
\end{align*}
approaches one for $n\to\infty$. By Lemma~\ref{lem:marginal}, joint typicality implies marginal typicality, so also $\Pr[(b_i^n,Y^n)\in\sett_{\epsilon_2}^n(P_{B_i Y})|\vecU^n(w,v)=\vecb^n]$ approaches one for $i=1,2,\dotsc,m$. This shows that $\Pr[\mathcal{E}_2]\overset{n\to\infty}{\to}0$.

%%%
\emph{Third error event:} 
By \eqref{eq:cond 2}, we have
\begin{align}
|\sett^n_{\epsilon_2}(P_{B_iY}|y^n)|\leq 2^{n\entop(B_i|Y)(1+\epsilon_2)}.
\end{align}
The size of $\hat{\setb}(y^n)$ is thus bounded as
\begin{align}
|\hat{\setb}(y^n)|\leq 2^{n\sum_{i=1}^m \entop(B_i|Y)(1+\epsilon_2)}.\label{eq:rate:sizeB}
\end{align}
Furthermore, by our random coding experiment, we have
\begin{align}
\Pr[\vecU^n(\tilde{w},\tilde{v})=\vecb^n]=2^{-nm}=2^{-n\entop(\vecU)},\quad\forall\vecb^n\in\{0,1\}^{mn}.\label{eq:uniprob}
\end{align}
We have
\begin{align}
\Pr[\mathcal{E}_3]&=\Pr\Bigl[\bigcup_{\substack{\tilde{w}\neq w\\\tilde{v}}}\vecU^n(\tilde{w},\tilde{v})\in\hat{\setb}(y^n)|Y^n=y^n\Bigr]\\
&=\Pr\Bigl[\bigcup_{\substack{\tilde{w}\neq w\\\tilde{v}}}\vecU^n(\tilde{w},\tilde{v})\in\hat{\setb}(y^n)\Bigr]\label{a3}\\
&\leq (2^{nR}-1)2^{n\tilde{R}}\Pr[\vecU^n\in\hat{\setb}(y^n)]\label{a4}\\
&< 2^{n(R+\tilde{R})}\Pr[\vecU^n\in\hat{\setb}(y^n)]\\
&= 2^{n(R+\tilde{R})}\sum_{\vecb^n\in\hat{\setb}(y^n)}\Pr[\vecU^n=\vecb^n]\\
&\overset{\eqref{eq:uniprob}}{=} 2^{n(R+\tilde{R})}\sum_{\vecb^n\in\hat{\setb}(y^n)}2^{-n\entop(\vecU)}\\
&\overset{\eqref{eq:rate:sizeB}}{\leq} 2^{n(R+\tilde{R})}2^{n\sum_{i=1}^m\entop(B_i|Y)(1+\epsilon_2)}2^{-n\entop(\vecU)}\label{eq:rate:ebound2}
\end{align}
where \tcr{equality in \eqref{a3}} follows because $\vecU^n(w,v)$ was transmitted, so $Y^n$ and $\vecU^n(\tilde{w},\tilde{v})$ are independent for $\tilde{w}\neq w$. \tcr{Inequality in \eqref{a4}} follows by the union bound. By \eqref{eq:rate:ebound2}, the probability $\Pr[\mathcal{E}_3]$ approaches zero for $n\to\infty$ if
\begin{align}
R+\tilde{R}+\sum_{i=1}^m\entop(B_i|Y)(1+\epsilon_2)-\entop(\vecU)<0.\label{eq:positive rate}
\end{align}
By choosing $\tilde{R}$ according to \eqref{eq:rate:cbmm}, condition \eqref{eq:positive rate} becomes
\begin{align}
R&<-\tilde{R}-\sum_{i=1}^m\entop(B_i|Y)(1+\epsilon_2)+\entop(\vecU)\\
&<-\sum_{i=1}^m\entop(B_i|Y)(1+\epsilon_2)+\entop(\vecU)-D(P_{\vecB}\Vert P_{\vecU})-\epsilon_1\log_2\frac{1}{\mu_{\vecB}\mu_{\vecU}}\\
&=\entop(\vecB)-\sum_{i=1}^m\entop(B_i|Y)(1+\epsilon_2)-\epsilon_1\log_2\frac{1}{\mu_{\vecB}\mu_{\vecU}}\label{eq:condition R}
\end{align}
where \tcr{equality in \eqref{eq:condition R}} follows because $\vecU$ is uniformly distributed, thus $D(P_{\vecB}\Vert P_{\vecU})=\entop(\vecU)-\entop(\vecB)$. Suppose now $\entop(\vecB)-\sum_{i=1}^m\entop(B_i|Y)>0$. Then, for any positive $R<\entop(\vecB)-\sum_{i=1}^m\entop(B_i|Y)$, we can find small enough positive $\epsilon_1<\epsilon_2$ so that both condition \eqref{eq:rate:cbmm} and \eqref{eq:positive rate} are fulfilled. By choosing $n$ large enough, the probability of the three error events approaches zero. If $\entop(\vecB)-\sum_{i=1}^m\entop(B_i|Y)\leq 0$, we let the transmitter transmit a dummy sequence corresponding to a rate of zero, which can be achieved on any channel. This shows that $\rbmd$ as defined in \eqref{rate} can be achieved by BMD.
%%%
\color{black}

\section{Proof of Lemma~\ref{lem:lm}}
\label{app:gmi}
We have
\begin{align}
\tcr{\rate(P_{\vecB},q_\textnormal{BMD},s_\textnormal{BMD},r_\textnormal{BMD})}=&\underbrace{\expop\left[\log_2\frac{\prod_{i=1}^m P_{B_i}(B_i)}{P_{\vecB}(\vecB)}\right]}_{=\entop(\vecB)-\sum_{i=1}^m\entop(B_i)}+\underbrace{\expop\left[\log_2\prod_{i=1}^mp_{Y|B_i}(Y|B_i)\right]}_{=-\sum_{i=1}^m\dentop(Y|B_i)}\nonumber\\
&\qquad-\underbrace{\expop\left[\log_2\left(\sum_{\vecb\in\supp P_{\vecB}}P_{\vecB}(\vecb)\frac{\prod_{i=1}^m P_{B_i}(b_i)}{P_{\vecB}(\vecb)}\prod_{j=1}^mp_{Y|B_j}(Y|b_j)\right)\right]}_{(\star)}\label{eq:gmiterms}
\end{align}
\tcr{where $\dentop(\cdot)$ denotes differential entropy in bits.} For the term $(\star)$, we have
\begin{align}
(\star)&=\expop\left[\log_2\left(\sum_{\vecb\in\supp P_{\vecB}}\prod_{i=1}^m P_{B_i}(b_i)p_{Y|B_i}(Y|b_i)\right)\right]\\
&\leq\expop\left[\log_2\left(\sum_{\vecb\in\{0,1\}^m}\prod_{i=1}^m P_{B_i}(b_i)p_{Y|B_i}(Y|b_i)\right)\right]\label{a2}\\
&=\expop\left[\log_2\prod_{i=1}^m\Bigl(\sum_{b\in\{0,1\}}P_{B_i}(b)p_{Y|B_i}(Y|b)\Bigr)\right]\\
&=\expop\left[\log_2\prod_{i=1}^mp_Y(Y)\right]\\
&=-\sum_{i=1}^m\dentop(Y)\label{eq:gmipy}
\end{align}
\tcr{with equality in \eqref{a2} if and only if $P_{\vecB}$ is strictly positive}. Using \eqref{eq:gmipy} in \eqref{eq:gmiterms}, we have
\begin{align}
\tcr{\rate(P_{\vecB},q_\textnormal{BMD},s_\textnormal{BMD},r_\textnormal{BMD})}&\geq\entop(\vecB)-\sum_{i=1}^m\entop(B_i)+\sum_{i=1}^m\bigl[\dentop(Y)-\dentop(Y|B_i)\bigr]\label{a1}\\
&=\entop(\vecB)-\sum_{i=1}^m\entop(B_i)+\sum_{i=1}^m\miop(B_i;Y)\label{eq:proof:bmdgmi}\\
&=\entop(\vecB)-\sum_{i=1}^m\entop(B_i)+\sum_{i=1}^m\bigl[\entop(B_i)-\entop(B_i|Y)\bigr]\\
&=\entop(\vecB)-\sum_{i=1}^m\entop(B_i|Y)
\end{align}
\tcr{with equality in \eqref{a1} if and only if $P_{\vecB}$ is strictly positive}.
\bibliographystyle{IEEEtran}
\normalsize
\bibliography{IEEEabrv,confs-jrnls,references}

\begin{thebibliography}{10}
\providecommand{\url}[1]{#1}
\csname url@samestyle\endcsname
\providecommand{\newblock}{\relax}
\providecommand{\bibinfo}[2]{#2}
\providecommand{\BIBentrySTDinterwordspacing}{\spaceskip=0pt\relax}
\providecommand{\BIBentryALTinterwordstretchfactor}{4}
\providecommand{\BIBentryALTinterwordspacing}{\spaceskip=\fontdimen2\font plus
\BIBentryALTinterwordstretchfactor\fontdimen3\font minus
  \fontdimen4\font\relax}
\providecommand{\BIBforeignlanguage}[2]{{%
\expandafter\ifx\csname l@#1\endcsname\relax
\typeout{** WARNING: IEEEtran.bst: No hyphenation pattern has been}%
\typeout{** loaded for the language `#1'. Using the pattern for}%
\typeout{** the default language instead.}%
\else
\language=\csname l@#1\endcsname
\fi
#2}}
\providecommand{\BIBdecl}{\relax}
\BIBdecl

\bibitem{bocherer2014probabilistic}
G.~B{\"o}cherer, ``Probabilistic signal shaping for bit-metric decoding,'' in
  \emph{Proc. IEEE Int. Symp. Inf. Theory (ISIT)}, Honolulu, HI, USA, Jun.
  2014, pp. 431--435.

\bibitem{zehavi1992psk}
E.~Zehavi, ``8-{PSK} trellis codes for a {R}ayleigh channel,'' \emph{{IEEE}
  Trans. Commun.}, vol.~40, no.~5, pp. 873--884, May 1992.

\bibitem{caire1998bit}
G.~Caire, G.~Taricco, and E.~Biglieri, ``Bit-interleaved coded modulation,''
  \emph{{IEEE} Trans. Inf. Theory}, vol.~44, no.~3, pp. 927--946, May 1998.

\bibitem{martinez2009bit}
A.~Martinez, A.~Guill{\'e}n~i F{\`a}bregas, G.~Caire, and F.~Willems,
  ``Bit-interleaved coded modulation revisited: A mismatched decoding
  perspective,'' \emph{{IEEE} Trans. Inf. Theory}, vol.~55, no.~6, pp.
  2756--2765, Jun. 2009.

\bibitem{ifabregas2010bit}
A.~Guill\'en~i F\`abregas and A.~Martinez, ``Bit-interleaved coded modulation
  with shaping,'' in \emph{Proc. IEEE Inf. Theory Workshop (ITW)}, Dublin,
  Ireland, Aug. 2010, pp. 1--5.

\bibitem{kaplan1993information}
G.~Kaplan and S.~Shamai~(Shitz), ``Information rates and error exponents of
  compound channels with application to antipodal signaling in a fading
  environment,'' \emph{AE\"U}, vol.~47, no.~4, pp. 228--239, 1993.

\bibitem{peng2012fundamentals}
L.~Peng, ``Fundamentals of bit-interleaved coded modulation and reliable source
  transmission,'' Ph.D. dissertation, University of Cambridge, 2012.

\bibitem{ganti2000mismatched}
A.~Ganti, A.~Lapidoth, and E.~Telatar, ``Mismatched decoding revisited: General
  alphabets, channels with memory, and the wide-band limit,'' \emph{{IEEE}
  Trans. Inf. Theory}, vol.~46, no.~7, pp. 2315--2328, Nov. 2000.

\bibitem{peng2013improved}
L.~Peng, A.~Guill\'en~i F\`abregas, and A.~Martinez, ``Improved exponents and
  rates for bit-interleaved coded modulation,'' in \emph{Proc. IEEE Int. Symp.
  Inf. Theory (ISIT)}, Istanbul, Turkey, Jul. 2013, pp. 1989--1993.

\bibitem{steiner2016protograph}
F.~Steiner, G.~B\"ocherer, and G.~Liva, ``Protograph-based {LDPC} code design
  for shaped bit-metric decoding,'' \emph{{IEEE} J. Sel. Areas Commun.},
  vol.~34, no.~2, pp. 397--407, Feb. 2016.

\bibitem{buchali2016rate}
F.~Buchali, F.~Steiner, G.~B\"ocherer, L.~Schmalen, P.~Schulte, and W.~Idler,
  ``Rate adaptation and reach increase by probabilistically shaped 64-{QAM}: An
  experimental demonstration,'' \emph{J. Lightw. Technol.}, vol.~34, no.~7,
  Apr. 2016.

\bibitem{gallager1968information}
R.~G. Gallager, \emph{Information Theory and Reliable Communication}.\hskip 1em
  plus 0.5em minus 0.4em\relax John Wiley \& Sons, Inc., 1968.

\bibitem{bocherer2015bandwidth}
G.~B{\"o}cherer, F.~Steiner, and P.~Schulte, ``Bandwidth efficient and
  rate-matched low-density parity-check coded modulation,'' \emph{{IEEE} Trans.
  Commun.}, vol.~63, no.~12, pp. 4651--4665, Dec. 2015.

\bibitem{achtenberg2013theoretic}
\BIBentryALTinterwordspacing
S.~Achtenberg and D.~Raphaeli, ``Theoretic shaping bounds for single letter
  constraints and mismatched decoding,'' \emph{arXiv}, 2013. [Online].
  Available: \url{http://arxiv.org/abs/1308.5938v1}
\BIBentrySTDinterwordspacing

\bibitem{gray1953pulse}
F.~Gray, ``Pulse code communication,'' U. S. Patent 2\,632\,058, 1953.

\bibitem{blahut1972computation}
R.~Blahut, ``Computation of channel capacity and rate-distortion functions,''
  \emph{{IEEE} Trans. Inf. Theory}, vol.~18, no.~4, pp. 460--473, Jul. 1972.

\bibitem{arimoto1972algorithm}
S.~Arimoto, ``An algorithm for computing the capacity of arbitrary discrete
  memoryless channels,'' \emph{{IEEE} Trans. Inf. Theory}, vol.~18, no.~1, pp.
  14--20, Jan. 1972.

\bibitem{kschischang1993optimal}
F.~R. Kschischang and S.~Pasupathy, ``Optimal nonuniform signaling for
  {G}aussian channels,'' \emph{{IEEE} Trans. Inf. Theory}, vol.~39, no.~3, pp.
  913--929, May 1993.

\bibitem{alvarado2011high}
A.~Alvarado, F.~Br\"annstr\"om, and E.~Agrell, ``High {SNR} bounds for the
  {BICM} capacity,'' in \emph{Proc. IEEE Inf. Theory Workshop (ITW)}, Paraty,
  Brazil, Oct. 2011, pp. 360--364.

\bibitem{bocherer2012efficient}
G.~B\"ocherer, F.~Altenbach, A.~Alvarado, S.~Corroy, and R.~Mathar, ``An
  efficient algorithm to calculate {BICM} capacity,'' in \emph{Proc. IEEE Int.
  Symp. Inf. Theory (ISIT)}, Cambridge, MA, USA, Jul. 2012, pp. 309--313.

\bibitem{kramer2007topics}
G.~Kramer, ``Topics in multi-user information theory,'' \emph{Foundations and
  Trends in Comm. and Inf. Theory}, vol.~4, no. 4--5, pp. 265--444, 2007.

\bibitem{masseyapplied1}
\BIBentryALTinterwordspacing
J.~L. Massey, ``Applied digital information theory {I},'' lecture notes, {ETH
  Zurich}. [Online]. Available:
  \url{http://www.isiweb.ee.ethz.ch/archive/massey_scr/adit1.pdf}
\BIBentrySTDinterwordspacing

\bibitem{elgamal2011network}
A.~El~Gamal and Y.-H. Kim, \emph{Network Information Theory}.\hskip 1em plus
  0.5em minus 0.4em\relax Cambridge University Press, 2011.

\bibitem{orlitsky2001coding}
A.~Orlitsky and J.~R. Roche, ``Coding for computing,'' \emph{{IEEE} Trans. Inf.
  Theory}, vol.~47, no.~3, pp. 903--917, Mar. 2001.

\end{thebibliography}

\end{document}